\documentclass{article}

\usepackage{graphicx}          
\usepackage{dcolumn}           
\usepackage{bm}                
\usepackage{float}             

\title{The age of the universe, the Hubble constant, the accelerated 
expansion and the Hubble effect}

\author{Domingos Soares\footnote{\small dsoares@fisica.ufmg.br} \\ Departamento de F\'{\i}sica, 
ICEx, UFMG --- C.P. 702 \\ 30123-970,  Belo Horizonte --- Brazil} 

\date{\today}

\begin{document}

\def\Ho{$H_\circ$}
\def\omegao{$\Omega_\circ$}
\def\to{$t_\circ$}

\maketitle

\begin{abstract}
The idea of an accelerating universe comes almost simultaneously with the 
determination of Hubble's constant by one of the Hubble Space Telescope Key 
Projects. The age of the universe dilemma is probably the link between 
these two issues. In an appendix, I claim that ``Hubble's law" might yet to be investigated 
for its ultimate cause, and suggest the ``Hubble effect" as  the searched candidate.
\end{abstract}

\bigskip
\bigskip

\section{The age dilemma}
The age of the universe is calculated by two different ways. Firstly, 
a lower limit is given by the age of the presumably oldest objects 
in the Milky Way, e.g., globular clusters. Their ages are calculated 
with the aid of stellar evolution models which yield 14 Gyr and 10\% 
uncertainty. These are fairly confident figures since the basics of stellar 
evolution are quite solid. Secondly, a cosmological age based on the 
Standard Cosmology Model derived from the Theory of General Relativity. 
The three basic models of relativistic cosmology are given by the 
Friedmann's solutions of Einstein's field equations. The models are 
characterized by a decelerated expansion from a spatial singularity 
at cosmic time $t=0$, and whose magnitude is quantified by  
the density parameter \omegao, the present ratio of the mass 
density in the universe to the so-called critical mass density. 
The critical Friedmann model has the critical mass density, and therefore, 
it has \omegao=1, which implies a spatially flat geometry. 
The present {\it observed} density parameter of the universe 
is approximately \omegao = 0.01, all made of baryonic matter, the 
usual matter in stars, planets and human beings. But the total mass density 
parameter ---  baryonic plus non baryonic, visible and dark --- is 
$\Omega_{m\circ} = 0.3$, derived from large-scale structure 
dynamics, Given that the non-critical 
Friedmann models are highly unstable at time $t=0$, meaning that any 
minute difference from a critical model would result, at time 
\emph{t = \to\ (now)}, an immensely large difference from \omegao = 1, 
it is generally accepted that the density 
parameter is precisely equal to 1. The discrepancy with the observed \omegao\ 
is considered as circumstantial evidence of the incompleteness nature 
of science. Eventually, one should find the reason for the difference. 
The \emph{fiducial} cosmological age of the universe is thus naturally 
given by the age of the critical model. It amounts to \to = $(2/3)H^{-1}_\circ$, 
with \Ho\ being the present Hubble constant. Or, \to = 6.5 $h^{-1}$ 
Gyr ($h$ being the Hubble constant in units of 100 km s$^{-1}$ Mpc$^{-1}$). 
Before the 1990s, $h$ was rather uncertain, ranging from 0.5 to 1. 
The lowest $h$ puts the fiducial cosmological age at 
acceptable --- yet marginal --- agreement with the stellar 
evolution age: \to = 13 Gyr. 

Then, two important projects in observational astronomy began in the early 1990s. 
The \emph{Hubble Space Telescope Key Project to Measure the Hubble Constant} 
and the efforts on using SNe Ia as standard candles to measure the cosmological 
deceleration. In 1995, Perlmutter et al. publish their result on one supernova 
at redshift z=0.458 and conclude that the universe is indeed decelerating at 
that cosmic epoch. Some years later, Madore et al. (1998) publish their 
first result on the HST Key Project: \Ho\ is in the range 
70 to 73 km s$^{-1}$ Mpc$^{-1}$. 
The fiducial cosmological age sits now in an uncomfortable narrower band, 
namely, 8.9--9.3 Gyr, inconsistent with the stellar evolution age.
But, in the next year, Perlmutter et al. publish their new results with an extended 
sample of SNe. They find now that the universe is in an accelerated expansion. 
The Friedmann critical model, modified to include a cosmological constant 
that drives the accelerated expansion, implies an age \to 
$= (2/3)H^{-1}_\circ\Omega_{\Lambda\circ}^{-1/2}
\ln[(1 - \Omega_{\Lambda\circ})^{-1/2}(1 + \Omega_{\Lambda\circ}^{1/2})]  
\cong$ (2/3)$H^{-1}_\circ(1-\Omega_{\Lambda\circ})^{-0.3}$, where $\Omega_{\Lambda\circ}$ 
is the density parameter associated to the cosmological constant 
(see Figure 1). With 
$\Omega_{m\circ} + \Omega_{\Lambda\circ} = 1$, and $\Omega_{m\circ} = 0.3$, 
one has now \to = 12.8--13.3 Gyr, again matching stellar ages. 

The age dilemma seems to come to an end. The solution is substantiated with two further 
amendments. Freedman et al. (2001) published the final results of the HST Key 
Project on \Ho. They confirm the previous range with the value \Ho\ $=72$  
km s$^{-1}$ Mpc$^{-1}$, and  10\% uncertainty as initially aimed. Riess et al. 
(2004), with an enlarged SNe Ia sample, the so-called \emph{Gold Sample}, 
confirmed the accelerating universe, adding the new finding that a transition 
from a decelerated phase occurs at a redshift of $0.46\pm0.13$, which matches 
Perlmutter's 1995 sole SN with extraordinary precision.

\begin{figure}[H]
\begin{center}
\includegraphics[width=7cm]{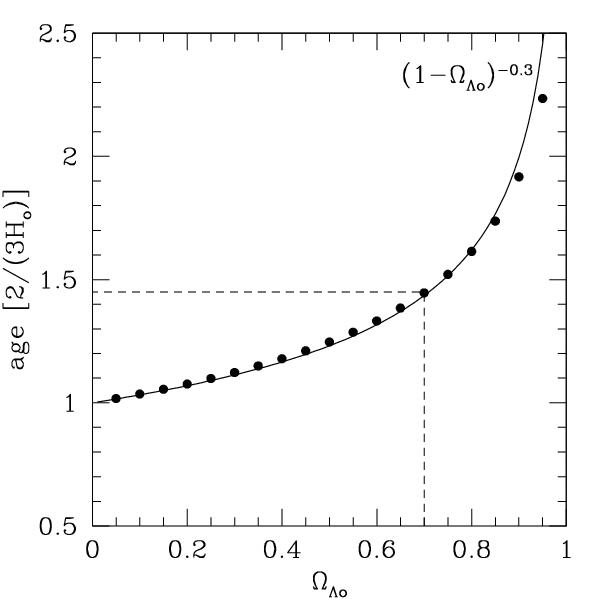}

\includegraphics[width=7cm]{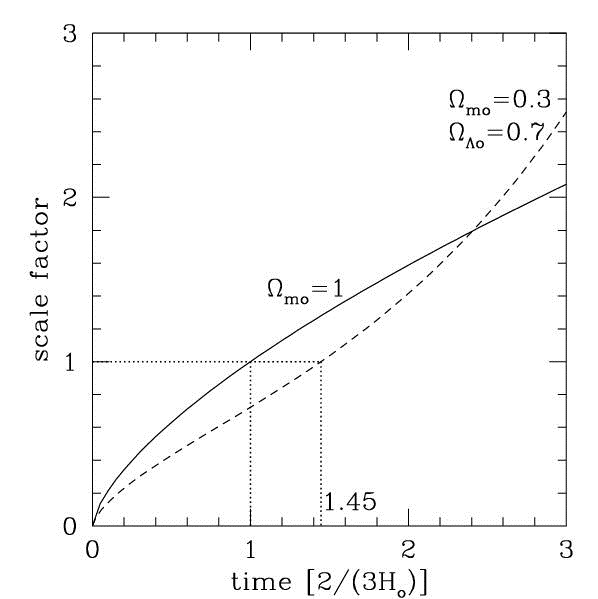}
\end{center}
\caption{Top panel: the age of the universe in an accelerated model as a 
function of $\Omega_{\Lambda\circ}$. Filled circles are the solution of 
Friedmann's equation with cosmological constant and the solid curve is an 
approximation to the exact solution.
Bottom panel shows the scale factor 
for the decelerated Einstein-de Sitter universe (solid curve) and for an universe 
with an accelerated phase at recent epochs (dashed curve). The age 
of the universe in both models is shown and corresponds to the scale factor of 
unity. Notice the changing of concavity of the dashed 
curve just before the scale factor of 1, meaning that the expansion 
has changed from a decelerated to an accelerated phase.}
\end{figure}

\section{Final remarks}
One big problem posed to the Standard Model is solved but others are 
raised. The existence and identification of the so-called {\it dark 
energy} is the most significant. Dark energy --- a generic name 
for what might be a cosmological constant or other candidates --- 
constitutes approximately 70\% of the mass-energy 
content of the universe and is responsible for its accelerated expansion.

Do you believe the accelerated expansion? Many do. Some do not.
One amongst the latter is John Archibald Wheeler (1911-2008). 
And he gives two reasons (Taylor and Wheeler 2000, p. G-11): 
{\it ``(1) Because the speed-up argument relies too trustingly 
on the supernovas being standard candles. (2) Because such an 
expansion would, it seems to me, contradict a view of 
cosmology too simple to be wrong."} Wheeler's second reason 
comes from his preferred cosmological model, namely, a closed 
Friedmann model (see p. G-1). He optimistically goes on saying 
that {\it ``Such clashes between theory and experiment have 
often triggered decisive advances in physics. We can hope that 
some decisive advance is in the offing.'' }

\bigskip

\bigskip

\appendix{\noindent\bf\Large Appendix}

\section{The Hubble effect}
``Hubble's law" is the linear relation between the logarithms of redshifts 
of distant cosmic bodies and their apparent magnitudes. There is not yet 
a satisfactory physical explanation to this law. I propose that the ``Hubble 
effect" --- yet to be elucidate --- be the physical mechanism responsible for 
Hubble's law. Accordingly, I put forward a tentative scientific guide for 
the discovery of such an effect. In this context, it turns out to be very 
useful a discussion of the heuristic description of the photoelectric effect 
made by Albert Einstein. \\~\\

\subsection{Introduction}
The American astronomer Edwin Powell Hubble (1889-1953) systematized and 
concluded the research about redshifts of distant galaxies, which was 
realized by himself and several astronomers during the first decades of 
the 20th century. 
The empirical relation between galaxy redshifts and their apparent 
brightness became known as ``Hubble's law". Such a relation is applied 
to a certain class of objects, namely, those that have the same intrinsic 
luminosity. For example, Hubble's law for the \emph{brightest galaxies of 
clusters} is very well illustrated in Fig. 2. There, one sees photographs  
of the brightest galaxies of five clusters of galaxies side by side with 
their spectra. 

\begin{figure}[H]
\begin{center}
\includegraphics[width=12cm]{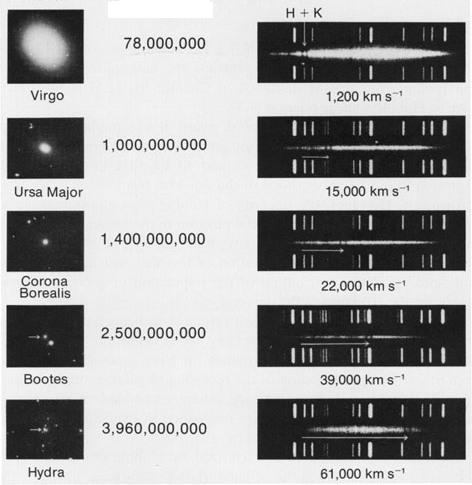}
\end{center}
\caption{\small Photographs, distances and redshifts for the 
brightest galaxies of clusters of galaxies. Galaxies are identified by 
the names of the clusters where they sit in. Distances, in the middle 
column, are given in light-years. Arrows, in the central section of 
each spectrum, indicate the shifts of lines H ($\lambda=$3968 \AA) 
and  K ($\lambda=$3934 \AA) of the chemical element calcium --- in its 
first ionization state, or CaII (``calcium two", in the astronomical 
jargon). The shifts are measured with respect to the same spectral lines 
observed from a stationary source in the laboratory. 
Redshifts $z=\Delta\lambda/\lambda$ 
are indicated below each spectrum as velocities, calculated with the 
expression of the non-relativistic Doppler shift $v=cz$, where $c$ is 
the speed of light in vacuum (Figure: Palomar Observatory, 
California Institute of Technology, United States). }
\end{figure}

The observational data available from Fig. 2 are redshifts $z$ --- 
obtained straight from the spectra --- and the apparent magnitudes 
of the galaxies --- obtained straight from the photographs. 
 
Assuming that all of the brightest galaxies of clusters have the same 
intrinsic luminosity, i.e., the same absolute magnitude, their distances 
can be calculated. The data shown in Fig. 2 are consistent with a 
direct proportionality between ``velocities" and distances $r$, that is, 
Hubble's law:  
$$ 
v=H_\circ r,    \eqno(1) 
$$
where the constant of proportionality $H_\circ$ is called ``Hubble's 
constant". Such an expression, nevertheless, is not the one which  
relates the observational data. The observational expression of Hubble's 
law is given by 
$$
m = 5\log cz + {\rm ~constant}, \eqno(2)
$$
where $m$ is the apparent magnitude (see Hoyle et al. 2000, Figs. 3.1 
and 3.2, Eq. 3.1, pp. 20-22).

The linear relation given by Eq. 1 was presented by Hubble for the 
last time in 1953 (Hubble 1953), the same year that he passed away. 
In the article, there is data from other clusters, besides those shown 
in Fig. 2, but the farthest galaxy is still the galaxy in the {\it Hydra} 
cluster, with $z=0.20.$ 

Hubble adopted the conversion of $z$ into velocity through the 
formula of the classical Doppler shift, but he considered it as an  
``apparent" velocity. In other words, he did not really believe that 
the redshift of a galaxy was caused by its motion of recession from 
us, which would justify the use of the Doppler shift expression.  
Contrary to what is presented in textbooks and in popular presentations 
of cosmology, Hubble did not accepted the idea of an expanding universe.
He had many reasons for that --- all of them motivated by his observational 
work. A detailed discussion of such a characteristic of Hubble's thought 
was made by Assis et al. (2008).   

One of the most strong reasons considered by Hubble was the question of 
the age of the universe. For an expanding universe, its age is approximately 
given by the inverse of Hubble's constant --- note that Hubble's constant 
has the physical dimension of the inverse of time. Using the value of 
Hubble's constant known up to the year of his death, 1953, the age of the 
universe resulted smaller than the geological age of Earth. That was a 
sign that Hubble, a scientist extremely aware of the value of 
observational and experimental data, could not ignore. 

\subsection{Redshift and velocity}
The Standard Model of Cosmology (SMC) --- also known as the Big-Bang 
model --- gives the function $v(z)$, which can be inserted in Eq. 2.

In what follows, I show how the SMC transforms Eq. 2 into the popular 
equation $v=H_\circ r$ (Eq. 1).

As an example of the expected result from the SMC, I shall use the 
critical Friedmann model (flat spatial geometry; see de Souza 2004, chaps. 
2 and 3, and Harrison 2000, chaps. 14 and 15). Redshift is caused by the 
expansion of space. The space-time in relativistic theories, i.e., derived 
from the Theory of General Relativity, is characterized by a metric that 
can vary with time. Thus, in Friedmann's models the space expands. Such an 
expansion is an intrinsic property of the space itself, not related 
to anything external. The value of $z$ depends on the expansion velocity, 
as illustrated in Fig. 3. It is worthwhile mentioning that the expansion 
velocity can be larger than the speed of light in vacuum. 

\begin{figure}[H]
\begin{center}
\includegraphics[width=8cm]{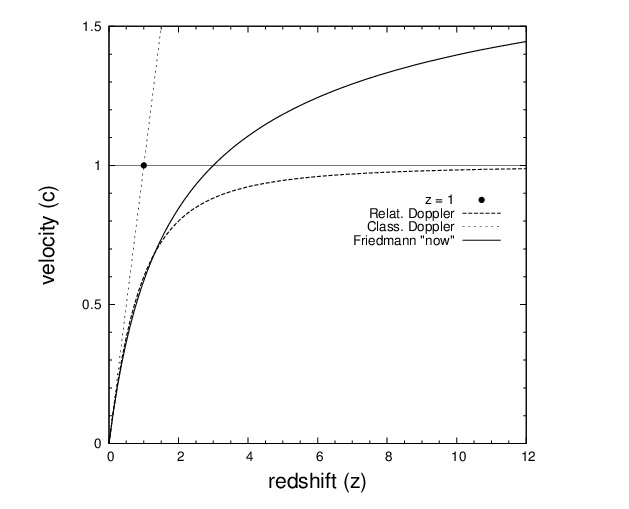}
\end{center}
\caption{\small Functions $v(z)$ for the Friedmann model and for the 
Doppler shift. The curve identified as {\it Friedmann ``now"} 
represents $v(z)$ corresponding to the expansion velocity of the universe 
prevalent at the instant of light detection. Note that the expansion 
velocity can be larger than the speed of light $c$. }
\end{figure}

For the Friedmann model, there are two possible functions: the observed 
value of $z$ can be associated to the expansion velocity of the universe 
at the instant of time in which the light was emitted, or to the 
expansion velocity at the instant in which the light was detected.   
Fig. 3 shows the last one, identified as {\it Friedmann ``now"} 
(see this same curve in Fig. 1 of Bedran 2002, where a comparison 
between the relativistic Doppler and cosmological redshifts is made).   
  
Fig. 3 shows also, for comparison, the expected functions $v(z)$ from the 
Doppler effect --- classical and relativistic. In the Doppler effect, the 
redshift is caused by the motion of recession of the light source. 

\begin{figure}
\begin{center}
\includegraphics[width=8cm]{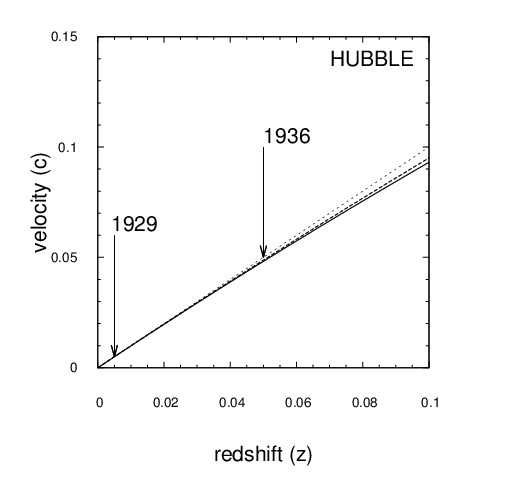}
\end{center}
\caption{\small Functions $v(z)$ of Fig. 3 are shown in a restrict 
range of redshifts, $0\le z\le 0.1$. Arrows indicate the maximum values 
of $z$ presented by Hubble in his articles of 1929 and 1936. Note that 
the three functions do not substantially differ in that range of redshifts.}
\end{figure}

Hubble presented, for the first time, the relation given by Eq. 1 in 1929 
(Hubble 1929), and later increased the range of redshifts (Hubble 1936). 
The article of 1929 is generally considered as the discovery article 
of Hubble's law, although recently there has been much dispute over this 
(e.g., Livio 2011 and references therein).  

Fig. 4 shows the same curves of Fig. 3 but now for the interval 
$0\le z \le 0.1$. In this range of $z$, all functions $v(z)$ can be 
written as:
$$
v(z) \simeq cz. \eqno(3)
$$
In the case of the classical Doppler shift we have $v(z)=cz$ exactly. 

Hubble's law is only valid for redshifts much smaller than unity 
($z\ll 1$), i.e., those considered in Fig. 4. In this range of $z$, the 
Friedmann model may, therefore, be approximated by $v(z)=cz$. 

In order to get Eq. 1 from Eq. 2, we have to use the approximation 
$v(z)=cz$ and the definition of apparent magnitude $m$. In terms of the 
absolute magnitude $M$ and of the distance $r$, the apparent magnitude is 
given by:
$$
m-M=5\log r-5, {\rm ~or} 
$$
$$
m=5\log r + {\rm ~constant}.  \eqno(4)
$$
By inserting Eq. 4 and $v(z)=cz$ in Eq. 2, we get $v=H_\circ r$. Consequently  
it becomes evident that Hubble's law is {\it consistent} with a model of 
expanding space. 

The trouble is that the model of the expanding universe, described by 
the SMC, {\it is not proved by the observations}. In order that such a 
model be valid it is necessary to admit the existence of substantial 
quantities of {\it dark matter} --- that is, undetected matter --- and of 
a {\it dark} component of non electromagnetic energy. Only 0.5\% of the 
total content of mass and energy of the universe are directly observed; 
a summary of the amounts of matter and energy in the universe, according to 
the SMC, is described in Soares (2002). {\it Dark energy} has non trivial 
properties and would be responsible for the accelerated expansion of the 
universe, prevailing at the current cosmic epoch. The reality of the 
accelerated expansion is also questionable, as discussed in the main text.   

In conclusion, the SMC is not completely satisfactory for the explanation of 
Hubble's law, as seen above. It turns out to be perfectly reasonable, then, 
the search for a physical mechanism that offers an alternative to that  
explanation. 

\subsection{A physical mechanism for Hubble's law}
Hubble's observations are consistent with the idea of an expanding 
universe, but they are not necessarily a proof of it. Hubble himself was 
aware of that and searched during all his life the correct answer for the 
question raised by his discovery: what does cause redshifts? (Assis 
et al. 2008).  

A possibility considered by him was the so-called {\it tired-light 
paradigm}. This was originally conceived by one of the greatest friends 
of Hubble's, Fritz Zwicky (Zwicky 1929; Soares 2014). We shall seek, therefore, 
a physical mechanism --- the {\it Hubble effect} --- valid for the tired-light 
paradigm. Generally speaking, the tired-light paradigm states that light 
looses energy --- its wavelength increases ---, when it ``travels" from the 
source to the observer. 
    
What physical mechanism could be this one? 

At this point, it is worthwhile remembering what happened in the past, 
in a similar situation, when Einstein put forward a heuristic 
interpretation of the {\it photoelectric effect}. One can make 
here a very useful counterpoint to the path for the discovery of the 
mechanism responsible for the Hubble effect.  

Einstein's heuristic model was based on the following experimental 
evidences (see discussion in Stachel 1998, p. 36): 

\begin{description}
\item{(a)} the effect does not depend on the intensity of the source of 
radiation; 
\item{(b)} the blackbody radiation, for short wavelengths, is described 
by Wien's limit; 
\item{(c)} the blackbody radiation, for large wavelengths, is described 
by the Rayleigh-Jeans distribution.
\end{description}

The items (b) and (c) were incorporated in Planck's explanation of the 
blackbody radiation, which also introduced  the {\it quanta} of energy 
in physics. Such a conception led Einstein to the idea that light consists 
of quanta of energy, and from there to the explanation of the photoelectric 
effect (Stachel 1998, p. 217). However, in 1905, Einstein did not use 
the complete distribution law of radiation for the blackbody, obtained 
by Planck in 1900, to the formulation of his model for the photoelectric effect 
(cf. Stachel 1998, p. 37).  

A heuristic program for the Hubble effect might, in a similar manner, 
contemplate the following observational evidences:

\begin{description}
\item{(a)} the effect depends on the flux of the source of radiation according 
to Hubble's law, given by Eq. 2; 
\item{(b)} the redshift does not depend --- or, has a very weak dependence --- 
on the wavelength of the radiation; 
\item{(c)} the effect is quantized (Tifft 2003, Arp 1998 and references 
therein);
\item{(d)} the effect does not cause light scattering which is, in general, selective, i.e., wavelength-dependent.  Otherwise, images of distant extended sources would be ``blurred", as if they were out of focus, which is not observed (cf. Harrison 2000, p. 312). 
\end{description}

In general, it is not observed large discrepancies in the redshift of a 
distant source, with respect to the measured wavelength (e.g., Sandage 1978 
and Rood 1982, where redshifts in the optical and radio wavelengths 
are studied). The fact that $z$ be independent of $\lambda$ is a 
characteristic both of the SMC --- $z$ is a consequence of the expansion of 
the space --- and of the hypothesis of the Doppler effect --- $z=v/c.$
One should not discard, nevertheless, the possibility that there is a small 
dependence of $z$ with $\lambda$, which would be an interesting feature 
of the putative Hubble effect. 

We have, therefore, a program that would certainly open up the way to a 
satisfactory physical theory for the tired-light paradigm, in other words, 
to the discovery of the Hubble effect.

\bigskip

\noindent{\bf Acknowledgment ---} Prof. Andr\'e K.T. Assis is gratefully 
acknowledged for the careful reading of the appendix, and for calling 
my attention to the possibility that redshifts may have some 
dependence on $\lambda$.

\end{document}